\documentclass[prb,twocolumn,showpacs,amsmath,amssymb]{revtex4}

\usepackage{graphicx}
\usepackage{dcolumn}
\usepackage{bm}

\begin{document}

\newcommand{\br}{{\bf r}}
\newcommand{\bk}{{\bf k}}
\newcommand{\bq}{{\bf q}}
\newcommand{\bp}{{\bf p}}

\newcommand{\eps}{\varepsilon}
\newcommand{\p}{\varphi}
\newcommand{\be}{\begin{eqnarray}}
\newcommand{\ee}{\end{eqnarray}}

\title{Temperature-dependent quantum  electron transport in 2D point contact}

\author{ T. V. Krishtop }
\author{ K. E. Nagaev }
\affiliation{Kotelnikov Institute of Radioengineering and Electronics,  Mokhovaya 11-7, Moscow, 125009 Russia}

\date{\today}

\begin{abstract}
We consider a transmission of electrons through a two-dimensional ballistic point contact in the low-conductance regime below the 0.7-anomaly. The scattering of electrons by  Friedel oscillations of charge density results in a contribution to the conductance proportional to the temperature. The sign of this linear term depends on the range of the electron-electron interaction and appears to be negative for the relevant experimental parameters.
\end{abstract}


\maketitle

\section{Introduction}

In recent years, the effects of electron-electron interaction on the conductance of low-dimensional ballistic contacts have attracted a considerable interest. This is mainly due to the attempts to explain the conductance plateaus at $0.7(2e^2/h)$ in quantum point contacts\cite{Thomas} and $0.5(2e^2/h)$ in quantum wires\cite{Reilly}. Usually these effects are explained by an existence of a localized state in the contact\cite{Flambaum, Spivak}. The mechanism of localization is not fully understood, and different scenarios of formation of such a  state\cite{Wingreen, Wang, Rejec} were proposed. One of them involves electron backscattering from the oscillations of the electron density in quantum point contacts\cite{Sablikov}.
This mechanism is inherent to any constriction, and the oscillations of electron density  are actually observed in  experiments\cite{Topinka}.

The formation of the plateaus in the gate-voltage dependence of the conductance is not the only possible effect of electron scattering by the Friedel oscillations in 2D systems. Recently, it was shown that scattering by Friedel oscillations in a two-dimensional conductor with impurities results in a strong temperature dependence of the conductivity\cite{Zala, Rudin}. It was also predicted that this scattering may give rise to a zero-bias anomaly of tunneling into the edge of a 2D electron gas\cite{Shekhtman95}.

\begin{figure}[t]
 \includegraphics[width=0.6\linewidth]{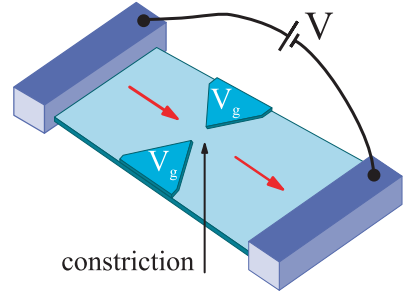}
 \caption{\label{fig1} Sketch of a realistic quantum point contact. The blue triangles show the gates that form the constriction, and the red arrows show the direction of motion of electrons.
 }
\end{figure}

Most of theoretical papers\cite{Flambaum, Wingreen, Wang, Rejec, Sablikov} dealt with interaction effects in the narrowest part of quantum point contacts (Fig. \ref{fig1}) by considering them as 1D channels and took into account only a few lowest transverse quantum modes. Therefore it is not clear how the transition to the continuum of quantum modes in the electrodes takes place and whether the interaction effects in the transition region outside the constriction  play a role. Meanwhile it is well known that Friedel oscillations in a 2D electron gas fall down with distance $x$ from a planar barrier according to the law $x^{-3/2}$, i.e. they penetrate deep into the electrodes. Hence their contribution to the conductance may be significant.

Recently, we considered  contacts much wider than the Fermi wavelength and obtained the temperature-dependent contribution to the conductance\cite{Nagaev08}, positive weak-field magnetoresistance\cite{Nagaev10}, and the shot noise\cite{Nagaev11} due to electron-electron interaction in the semiclassical approximation. The theoretically predicted $G(T)$ and $G(H)$ dependencies\cite{Nagaev08, Nagaev10} are in a good agreement with the experiments\cite{Renard, Khrapai}. However the linear in temperature contribution to the conductance persists even for contacts of the width approaching the Fermi wavelength\cite{Khrapai}, where the scattering by quantum oscillations of electron density should be considerable.
Therefore it is of interest to compare the contribution from this quantum effect with the semiclassical one.
To this end, we calculate the temperature-dependent correction to the conductance for the case of small contact size $a \ll \lambda_F$ and consider the interplay between the single-slit diffraction and the interaction. As we assume the interaction to be weak, the formation of a localized state in the contact is irrelevant to our problem.
We do not address here the physics related with the 0.7-anomaly and focus on the low-conductance regime.

The electron scattering by the Friedel oscillations results in a cusp in the probability of transmission through the contact at the Fermi surface, which leads to a linear temperature dependence of the conductance similarly to the semiclassical case. However the sign of this linear correction depends on the the competition between a negative contribution from the direct interaction and a positive contribution from the exchange interaction. The dependence of the absolute and relative corrections to the conductance on the contact size is also different from the semiclassical one.

The paper is organized as follows. In Section II, we present the model and describe our general formalism.  Section III addresses the case of noninteracting electrons, and Section IV describes the perturbation theory. Sections V and VI  present the results for a point-like interaction and a generalization for an interaction of a finite range, and Section VII contains the discussion of the results.

\section{General approach}

We consider the effects of electron-electron interaction on the conductance of a narrow short contact at non-zero temperature. We assume that electron-electron interaction is weak so that it can be treated perturbatively.

\begin{figure}[t]
 \includegraphics[width=0.6\linewidth]{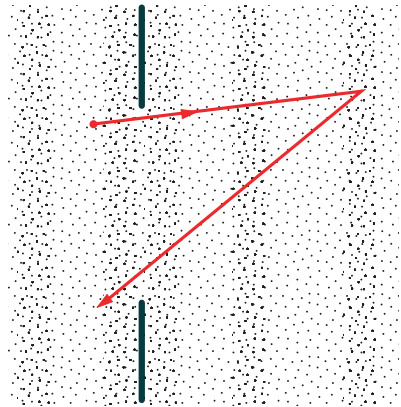}
 \caption{\label{fig2} The model of quantum point contact used in our calculations. Red line shows a process  of
  electron scattering  by the Friedel oscillations that affects the conductance.
 }
\end{figure}

As the Friedel oscillations die out at a large distance $v_F/T$ from an obstacle, we are mainly interested in scattering processes that occur in the regions outside the contact in the leads and do not focus on the exact dynamics of an electron in the narrowest part of the constriction.
Therefore we consider an extremely short contact, namely, we use typical single-slit diffraction model - a gap of width $2a \ll\lambda_F$ in an one-dimensional barrier separating two half-planes of 2DEG (Fig. \ref{fig2})\cite{Kawabata}. This model geometry allows us to avoid dealing with an infinite number of discrete transverse modes and to use instead the continuous representation.

Note that Friedel oscillations far from the barrier do not depend on the exact shape of the confinement potential because they are formed by electrons near the Fermi level with almost normal incidence on the barrier. Hence a smooth barrier potential should result only in a shift of their phase, which would not essentially change the correction to the conductance (see Appendix B).

We obtain the conductance by using a classical Landauer approach\cite{Landauer} and write the conductance as a sum of transmission coefficients

\be
G = g_s \frac{e^2}{\hbar} \int{\frac{d\eps}{2\pi}} \left(-\frac{\partial f}{\partial \eps}\right) \, \sum\limits_{\bk, \bq} |t(\bk, \bq)|^2.
\label{G}
\ee
Here $g_s$ is a spin degeneracy and $t(\bk, \bq)$ is the transmission amplitude from mode with the wave-vector $\bk$ in the left half-plane to the mode with the wave-vector $\bq$ to the right half-plane.

First of all we calculate  $t = t_0$ and $G = G_0$ for noninteracting electrons.
A weak electron-electron interaction results in a scattering of electrons by the Friedel oscillations caused by the contact boundaries. We consider the oscillations arising from the barrier as one-dimensional and neglect their distortion by the gap because this effect is of higher order in the contact size. The incident electron is scattered by the Friedel oscillations before and after passing through the contact, which results in a correction to the transmission coefficient of the contact $t(\bk, \bq) = t_0(\bk, \bq) + \delta t(\bk, \bq)$. The correction to the transmission coefficient may be obtained by expanding the perturbation of the wave function  $\delta \psi$ in plane waves. To calculate this perturbation, we solve a Schr\"odinger-type equation

\be
\left[-\frac{\hbar^2}{2m}\nabla^2 + V_{eff}(\br)\right]\psi(\br) = \eps\, \psi(\br)
\label{schrodinger}
\ee
in an iterative way\cite{Landau2} and obtain $\delta \psi(\br)$ in the lowest order in the interaction.

Now we discuss the interaction potential induced by the Friedel oscillations. It is a sum of a direct term and an exchange one\cite{Kittel} $V_{eff}(\br) = V_H(\br) - V_F(\br)$, where
\be
V_H (\br) = g_s \int{d\br_1 \, U_{ee}(\br - \br_1) \, n(\br_1, \br_1)},
\label{V_H = U n()}
\ee

\be
V_F (\br) \, \psi(\br) = \int{d\br_1 \, U_{ee}(\br - \br_1) \, n(\br, \br_1) \, \psi(\br_1)},
\label{V_F = U n()}
\ee
where
$
 n({\bf r}, {\bf r_1}) = \langle \hat\psi^{+}(\br_1)\,\hat\psi(\br)\rangle,
$
$\hat\psi^{+}$ and $\hat\psi$ are electron creation and annihilation operators, and $U_{ee}(\br - \br_1)$ is the potential of the electron-electron interaction. Typically, it is the Coulomb interaction screened by the two-dimensional electrons and by the gate.
The coefficient of spin degeneracy $g_s$ appears only in the direct term because it involves interactions between electrons with both spin directions while the exchange interaction is possible only for electrons with the same spin.

Equations (\ref{V_H = U n()}) and (\ref{V_F = U n()}) result in a correction to the wave function in the form $\delta \psi = \delta \psi_H - \delta \psi_F$ and, accordingly, in a correction to the conductance

\be
\delta G = \delta G_H - \delta G_F.
\ee
The negative sign in the exchange term is explicitly shown here.

\section{Non-interacting electrons}

In the absence of electron-electron interaction the conductance calculation reduces to the standard problem of diffraction by a narrow gap of width $a \ll \lambda_F$. For the three-dimensional case it was considered many times\cite{Landau} and the  conductance of a small three-dimensional ballistic contact\cite{Itskovich} was found to be proportional to the sixth power of the contact size $G \propto (k_Fa)^6$.

To the best of our knowledge, the two-dimensional problem in the limit of $a \ll \lambda_F $ was considered only once for a specifically designed model of the contact\cite{Zagoskin}.
In the limit of $k_F a \ll 1 $, these authors obtained $G \propto 1/\ln^2(k_Fa)$, which is unphysical. Therefore we recalculate this quantity using the solution of the problem of diffraction from a narrow slit obtained many decades ago in optics\cite{Sommerfeld}.

We use an approach similar to Sommerfeld\cite{Sommerfeld} and reduce the solution of the Schr\"odinger equation (\ref{schrodinger}) with
$V_{eff}=0$ to a boundary-value problem. The total wave function may be presented in the form
\be
\begin{cases}
\psi(\br') = \psi_0(\br') + \psi_t(\br'),& x' < 0 \\
\psi(\br') = \psi_t(\br'),& x' > 0,
\end{cases}
\label{psi = psi_0 + psi_t}
\ee
where $\psi_0$ is the wave function in the absence of the gap and $\psi_t$ is the lowest-order correction in the gap size. The zero-order wave function $\psi_0$ obeys zero boundary conditions both at the barrier and the gap, while the correction $\psi_t$ obeys the zero boundary condition at the barrier and a nonzero boundary condition at the gap
\be
\left\{
\begin{aligned}
&(\nabla^2 + k^2)\,\psi_t(x, y) = 0 \\
&\psi_t(x, y)|_{x = 0, \, y\in(-a, a)} = \chi(0, y),
\end{aligned}
\right.
\label{chi-system}
\ee
where $k^2 = 2m\eps$.

We assume that the incoming plane wave $\psi_i(\br) = \sqrt{m/k_x}\,e^{i (k_x x + k_y y)}$ with $k_x^2 + k_y^2 = k^2$ falls on the contact from the left in the $x$ direction and find the boundary condition $\chi(y)$ self-consistently  using the continuity of the  derivative of the total wave function at the gap (see Appendix A). Expanding the transmitted wave function in plane waves allows us to obtain transmission coefficient for noninteracting electrons
\be
t_0(\bk, \bq) = -\frac{i\pi}{2}a^2 \, \sqrt{k_x q_x}.
\label{t0}
\ee
We substitute it in Eq.(\ref{G}) and obtain
\be
G_0 = g_s \frac{e^2}{\hbar} \frac{\pi}{128} \, k_F^4 a^4 + {\cal O}\!\left(\frac{T^2}{E_F^2}\right).
\label{G0}
\ee
This contact-size dependence is more physically plausible than that of Ref. \onlinecite{Zagoskin} because it corresponds to the two-dimensional analog of the Rayleigh scattering of light by small particles\cite{Landau}. Indeed, the conductance is proportional to the square of the two-dimensional particle volume.

\section{Perturbation theory}

Now we take into account a weak electron-electron interaction. This interaction leads to a scattering of electrons by the Friedel oscillations induced by the barrier and results in a correction $\delta t(\bk, \bq)$ to the transmission coefficient. We substitute it in Landauer formula (\ref{G}) and obtain the correction to the conductance in the case of weak interaction
\begin{multline}
\delta G = -2g_s \frac{e^2}{\hbar} \int{\frac{d\eps}{2\pi}} \left(-\frac{\partial f}{\partial \eps}\right)
\\{}\times
 \sum\limits_{\bk, \bq} |t_0(\bk, \bq)| \,{\rm Im}\,\delta t(\bk, \bq).
\label{delta G = 2 t_0 delta t}
\end{multline}
Here we take into account the fact that $t_0$ (\ref{t0}) is an imaginary quantity.
The correction to the transmission coefficient is conveniently expressed in terms of the  correction to the wave function by expanding it in plane waves. The wave function is found by solving the Schr\"odinger equation (\ref{schrodinger}) in the lowest order in the interaction. To this end, we isolate the term with $V_{eff}$ in the right-hand side and  substitute the unperturbed wave function into it. The solution  is given by

\be
\delta \psi(\br) = \int{d\br' \, g(\br, \br') \, V_{eff}(\br') \, \psi(\br')}.
\label{delta psi = g V_eff psi}
\ee

Here $V_{eff}(\br')$ is the scattering potential produced by the Friedel oscillations, $g(\br, \br')$ and  $\psi(\br')$ are the single-electron Green function and the total wave function for noninteracting electrons. We assume that the electrons are incident on the contact from the left and we measure the total current on the right, where $x > 0$. We are interested in the entire range of values of $ x' \in (- \infty, \infty)$ because we consider the scattering by the Friedel oscillations on both sides of the contact.  Similarly to the wave function  (\ref{psi = psi_0 + psi_t}), the one-electron Green function may be written in the form
\be
\begin{cases}
g(\br, \br') = g_t(\br, \br'), & x' < 0 \\
g(\br, \br') = g_0(\br, \br') + g_t(\br, \br'), & x' > 0,
\end{cases}
\label{g = g_0 + gt}
\ee
where $g_0$ is the Green function in the absence of the gap, and $g_t \propto a^2$ is the second-order correction in the gap size. We calculate $g_t$ similarly to $\psi_t$ (see Appendix A) by solving the system
\be
\left\{
\begin{aligned}
&\frac{\hbar^2}{2m} \, (\nabla^2 + k^2)\, g(\br, \br') = \delta(\br - \br') \\
&g(\br, \br')|_{x = 0, \, y\in(-a, a)} = \chi(0, y, x', y').
\end{aligned}
\label{g-eq}
\right.
\ee
We substitute (\ref{psi = psi_0 + psi_t}) and (\ref{g = g_0 + gt}) into (\ref{delta psi = g V_eff psi}) and obtain in the lowest order in the contact size
\be
\delta \psi(\br) = \int\limits_{x' < 0}{d\br' \, g_t(\br, \br') \, V_{eff}(\br') \, \psi_0(\br')}
\nonumber\\ +
\int\limits_{x' > 0}{d\br' \, g_0(\br, \br') \, V_{eff}(\br') \, \psi_t(\br')}.
\label{delta psi = psi_0 g_t + g_0 psi_t}
\ee
The first term corresponds to electron scattering by Friedel oscillations in front of the contact, and the second one - behind it.

We substitute the expressions for the interaction potential in the Hartree-Fock approximation (\ref{V_H = U n()}) and (\ref{V_F = U n()}) into (\ref{delta psi = psi_0 g_t + g_0 psi_t}) and obtain the conductance as a sum of direct and exchange terms $\delta G = \delta G_H - \delta G_F$.
Then we substitute $\psi_0$, $\psi_t$, $g_0$, and $g_t$ into the resulting expression and after some simplifications obtain the conductance for an arbitrary interaction potential in the form
\begin{multline}
\delta G_H =  - g_s^2 \frac{e^2}{\hbar} \frac{m}{\hbar^2} \frac{1}{16} a^4  \int{d\eps}
\left(-\frac{\partial f}{\partial \eps}\right)k^2
\int\limits_{-\infty}^{\infty}{dy_1}
\int\limits_{0}^{\infty}{dx_1}
\\
\times
\int\limits_{0}^{\infty}{dx'}\, U_{ee}(x' - x_1, - y_1)\,n(x_1)
\int\limits_{-\infty}^{\infty}{dq_y}\,
\sin(2 q_x x')
\label{delta G_H}
\end{multline}

\begin{multline}
\delta G_F =  - g_s \frac{e^2}{\hbar} \frac{m}{\hbar^2}\frac{1}{8} a^4
\int{d\eps} \left(-\frac{\partial f}{\partial \eps}\right)k^2
\int\limits_{-\infty}^{\infty}{dy_1 }
\\
\times
\int\limits_{0}^{\infty}{dx_1}
\int\limits_{0}^{\infty}{dx'}\, U_{ee}(x' - x_1, - y_1)\,
n(\br', x_1, y_1 + y')
\\
\times
\int\limits_{-\infty}^{\infty}{dq_y}\,
\sin(q_x x')\,\cos(q_y y_1)\,\cos(q_x x_1)
\label{delta G_F}
\end{multline}
We use the coordinate transform $y_1 \to y_1 + y'$ to make the interaction potential independent of $y'$ and then integrate over $y'$. This transform results in the independence of the Friedel oscillations of density on $y'$ because we obtain the corrections in the lowest approximation in the contact size and use the unperturbed wave functions in the absence of the gap to calculate $n$ (see Appendix B).
\begin{multline}
n(\br', x_1, y_1 + y') = \frac{1}{2\pi}\int\limits_{0}^{\infty}{dp \left(-\frac{\partial f(p)}{\partial p}\right) p}
\\
\times
\left[\frac{J_1(p\sqrt{(x' - x_1)^2 + y_1^2})}{\sqrt{(x' - x_1)^2 + y_1^2}} - \frac{J_1(p\sqrt{(x' + x_1)^2 + y_1^2})}{\sqrt{(x' + x_1)^2 + y_1^2}}\right]
\label{friedel}
\end{multline}

By setting $x' = x_1$ and $y_1 = 0$ in this expression,  it is easy to obtain the electron density $n(x_1) = n(\br_1, \br_1)$ , which is responsible for the direct interaction term and depends only on one coordinate
\be
n(x_1) = \frac{k_F^2}{4\pi} - \frac{1}{2\pi}\int\limits_{0}^{\infty}dp \left(-\frac{\partial f}{\partial p}\right)
p\,
\frac{J_1(2px_1)}{2x_1}.
\label{n(x_1)}
\ee
The first term here presents a uniform charge density, and the second one describes its oscillations with a period
$(2k_F)^{-1}$ at large distances from the barrier that decay according to the law $x^{-3/2}$ at zero temperature. At nonzero temperature, they exponentially decay at a characteristic length $v_F/T$.

\section{Point-like interaction potential}

\begin{figure}[t]
 \includegraphics[width=0.6\linewidth]{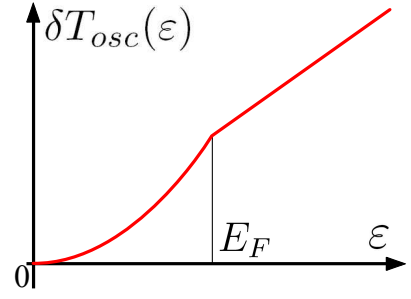}
 \caption{\label{fig3} Cusp in the transmission coefficient.
 }
\end{figure}

Consider now the case of a point-like interaction potential $U_{ee}(x' - x_1, - y_1) = U_p\,\delta(x' - x_1)\,\delta(y_1)$. A comparison of Eqs. (\ref{delta G_H}) and (\ref{delta G_F}) shows that $\delta G_H = g_s\, \delta G_F$. Therefore $\delta G = (g_s - 1)\, \delta G_F$. Upon an integration over $x_1$, $y_1$ and $q_y$, one obtains the correction in the form
\begin{multline}
\delta G = [1 - g_s]\,g_s \frac{e^2}{\hbar}\, \frac{m}{\hbar^2}\,\frac{\pi}{16}\, a^4 U_p
\\
\times
\int{d\eps} \left(-\frac{\partial f}{\partial \eps}\right)k^3
\int\limits_{0}^{\infty}{dx'}\, n(x')\, J_1(2kx').
\end{multline}
With $n(x')$ from (\ref{n(x_1)}) substituted into this expression, it is easily seen that the main contribution to it is given by values $x' \sim v_F/T$, i. e. by  the "tail" of the Friedel oscillations far from the barrier. We integrate over $x'$ and $p$ and calculate the total transmission coefficient $\delta T(\eps)$. It is a sum of two parts $\delta T_{const}(\eps) + \delta T_{osc}(\eps)$ formed by the constant and the oscillating part of the electron density (\ref{n(x_1)}), respectively. The term $\delta T_{const}(\eps)$ is a smooth function without singularities, whereas $\delta T_{osc}(\eps)$ has a cusp at the Fermi surface of the form
\be
 \delta T_{osc}(\eps)
 \propto
 \frac{\eps}{E_F}
 \left[
  \frac{\eps}{E_F} - \frac{T}{E_F}\,\ln\!\left({1 + e^{\frac{\eps - E_F}{T}}}\right)
 \right].
\ee
At $T/E_F \ll 1$, the derivative of the last term with respect to $\eps/E_F$ tends to 2 at $\eps = E_F - 0$ and to 1 at
$\eps = E_F + 0$.
This cusp (see Fig. \ref{fig3}) results in a linear temperature dependence of conductance
\be
\delta G = \frac{[1 - g_s]\,g_s}{128}\,
\frac{e^2}{\hbar}\, \frac{m}{\hbar^2}\, k_F^4 a^4 U_p
\frac{T}{E_F}.
\label{delta G point result}
\ee
Alternatively, this temperature dependence may be  attributed to the temperature-dependent cut-off length of the Friedel oscillations (see Fig. 1).

\section{Arbitrary interaction potential}

\subsection{Direct interaction}

An isotropic finite-range interaction potential is conveniently described by its Fourier components $U_p(p)$, which depend only on the absolute value of $\bp$.
Similarly to the case of a point interaction potential, the substitution of the two terms in (\ref{n(x_1)}) that correspond to the constant and the oscillating parts of the charge density  into (\ref{delta G_H}) results in a sum of two terms $\delta G_H = \delta G_{H, const} + \delta G_{H, osc}$. The first term is easily calculated and equals
\begin{multline}
\delta G_{H, const} =  - g_s^2 \frac{e^2}{\hbar} \frac{m}{\hbar^2} \frac{1}{128} k_F^4 a^4  U_p(0)
\\
+ {\cal O}\!
\left(e^{-E_F/T}\right)
\label{delta G_H_const_result}
\end{multline}
Here $U_p (0)$  is the Fourier transform of the interaction potential. After some simplifications, the second term may be brought to the form
\be
\delta G_{H, osc}=  g_s^2\, \frac{e^2}{\hbar}\, \frac{m}{\hbar^2}\, \frac{1}{128\pi}\, a^4
\int{dk}\left(-\frac{\partial f}{\partial k}\right)k^3
 \nonumber\\
\times
\int\limits_{0}^{\infty}dp \left(-\frac{\partial f}{\partial p}\right)p
\int{dp_{1}}\,
U_p(p_1)\,I_1(k, p, p_1),
\label{dG_H_osc}
\ee
where
we have introduced the notation
\begin{multline}
I_1 (p > k) =
\frac{\theta(-p_1 + 2k)}{4kp}\left[\sqrt{4p^2 - p_1^2} + \frac{p_1^2}{\sqrt{4k^2 - p_1^2}}\right]
\\
-
\frac{ \theta(-p_1 + 2p)\,\theta(p_1 - 2k) }{p_1 + \sqrt{p_1^2 - 4k^2}}\,
\frac{k}{p}\,\sqrt{ \frac{{4p^2 - p_1^2}}{{p_1^2 - 4k^2}} },
\end{multline}

\begin{multline}
I_1(p < k) =
\frac{\theta(-p_1 + 2p)}{4kp}\left[\sqrt{4p^2 - p_1^2} + \frac{p_1^2}{\sqrt{4k^2 - p_1^2}}\right]
\\
+
\frac{ \theta(p_1 - 2p)\,\theta(-p_1 + 2k) }
     { k \sqrt{4k^2 - p_1^2} }\,
\frac{ p\, p_1}{p_1 + \sqrt{p_1^2 - 4p^2}}.
\end{multline}
 The quantity  $I_1(k, p, p_1)$ has  singularities at $p_1 = 2k$ and $p_1=2p$,
and the  derivatives of the distribution function in (\ref{dG_H_osc}) cut out narrow intervals of $k$ and $p$ of width $\sim k_F T/E_F$ near $k_F$. As we assume the potential $U_p$ to be a smooth function of $p$ at the scale $T/v_F$, we can isolate the singular part of the integrand and substitute $U_p(p_1)=U_p(2k_F)$ in it, while setting $I_1(k, p, p_1) = I_1(k = k_F, p = k_F, p_1)$ in its regular part
\be
U_p(p_1)\,I_1(k, p, p_1) \approx U_p(2k_F)\,I_1(k, p, p_1)
+ \nonumber\\ +
[U_p(p_1) - U_p(2k_F)]\,I_1(k = k_F, p = k_F, p_1).
\ee
We calculate both terms and obtain the correction due to the direct interaction in the form
\be
\delta G_{H} =
- g_s^2\, \frac{e^2}{\hbar}\, \frac{m}{\hbar^2}\, \frac{1}{128}\, k_F^4 a^4\, U_p(2k_F)\,\frac{T}{E_F}
\nonumber\\ +
g_s^2\, \frac{e^2}{\hbar}\, \frac{m}{\hbar^2}\, \frac{1}{64 \pi}\, k_F^4 a^4
\int\limits_{0}^{2k_F}{dp_1}
\frac{U_p(p_1) - U_p(0)}{\sqrt{4k_F^2 - p_1^2}}.
\label{delta G_H result}
\ee
The first term here presents the contribution linear in temperature and is proportional to the Fourier component of the interaction potential at $2k_F$ while the second one presents the temperature-independent contribution and vanishes if $U_p(p)$ is a constant.

\subsection{Exchange interaction}

The substitution of the two terms of (\ref{friedel}) into (\ref{delta G_F}) gives the exchange contribution to the conductance in a form $\delta G_F = \delta G_{F, const} + \delta G_{F, osc}$ in  analogy with  $\delta G_H$.
The first term is easily calculated and equals
\begin{multline}
\delta G_{F, const} =  - g_s \frac{e^2}{\hbar}\, \frac{m}{\hbar^2}\,\frac{1}{64\pi}\, k_F^2 a^4
\\
\times
\int\limits_{0}^{2k_F}{dp_1}\,U_p(p_1)\,p_1\, \arccos\!\left(\frac{p_1}{2 k_F}\right)
+
{\cal O}\left(e^{-{E_F}/{T}}\right).
\label{delta G_F const result}
\end{multline}
After a simple rearrangement, the second term may be brought to the  form
\begin{multline}
\delta G_{F,osc} =  g_s \frac{e^2}{\hbar}\, \frac{m}{\hbar^2}\,\frac{1}{32 \pi^2}\, a^4
\int{dk \left(-\frac{\partial f}{\partial k} \right)k^2 }
\\ \times
\int\limits_{0}^{\infty}{dp \left(-\frac{\partial f}{\partial p}\right) }
\int\limits_{0}^{p}{dp_y}
\int\limits_{0}^{k}{dq_y}\,
U_p(p_y - q_y)\,
\ln\!\left|\frac{q_x + p_x}{q_x - p_x}\right|.
\label{delta_G_F_osc-1}
\end{multline}
This term has a singularity at $p_x = q_x$ because the backscattering of electrons is most efficient if the $x$ component of the electron momentum $q_x$ coincides with the wave vector $p_x$ of the Friedel oscillations. We write the integrand as a sum of two terms, one of which has a singularity at $p_x = q_x$ and the second one is a regular function, so that one may set $k = p = k_F$ in it to obtain
\be
U_p(p_y - q_y)\,\ln\left|\frac{q_x + p_x}{q_x - p_x}\right| \approx
U_p(0)\, \ln\left|\frac{q_x + p_x}{q_x - p_x}\right| \nonumber\\
+ \left.
[U_p(p_y - q_y) - U_p(0)]\,\ln\left|\frac{q_x + p_x}{q_x - p_x}\right|\right|_{p = k = k_F}.
\ee

We perform the integration in (\ref{delta_G_F_osc-1}), sum the result with (\ref{delta G_F const result}) and obtain the correction due to an exchange interaction in the form
\be
\delta G_{F} =
- g_s\, \frac{e^2}{\hbar}\, \frac{m}{\hbar^2}\,\frac{1}{128}\, k_F^4 a^4\, U_p(0)\, \frac{T}{E_F}
+ \delta G_{F, T=0},
\label{delta G_F result}
\ee
where $\delta G_{F, T=0}$ is a temperature-independent quantity given by an integral
\begin{multline}
\delta G_{F, T=0} = g_s\, \frac{e^2}{\hbar} \frac{m}{\hbar^2}\frac{1}{32\pi^2} k_F^3 a^4
\int\limits_{0}^{2k_F}{dp_1}\,U_{ee}(p_1)
\\
\times
\Biggl[
 K\left(\sqrt{1 - \frac{p^2_1}{4k_F^2}}\right) - E\left(\sqrt{1 - \frac{p^2_1}{4k_F^2}}\right)
\\
 - {\pi}\,\frac{p_1}{2k_F}\,\arccos\!\left(\frac{p_1}{2 k_F}\right)
\Biggr]
\label{dG_T=0_exch}
\end{multline}
where $K$ and $E$ are full elliptic integrals of the first and second kind. The temperature-dependent correction to the conductance in (\ref{delta G_F result}) is determined by the long-wavelength component of the interaction potential, which is typical for the exchange interaction\cite{Altshuler,Zala}. However $\delta G_{F, T=0}$ is determined by all components of $U_p$ from 0 to $2k_F$. Long-wavelength components contribute to (\ref{dG_T=0_exch}) with a positive sign and short-wavelength components contribute to it with negative sign, so that the integral is zero if $U_p$ is constant.

\section{Discussion}

The summation of (\ref{G0}), (\ref{delta G_H result}) and (\ref{delta G_F result}) gives the full conductance $G = G_0 + \delta G_H - \delta G_F$ in the form
\be
G = g_s\, \frac{e^2}{\hbar}\, \frac{\pi}{128} \, k_F^4 a^4 + \delta G_{T = 0}
\nonumber\\ +
g_s\, \frac{e^2}{\hbar}\, \frac{m}{\hbar^2}\, \frac{1}{128}\, k_F^4 a^4\, [U_p(0) - g_s U_p(2k_F)]\,\frac{T}{E_F}.
\label{G_0+G_H-G_F}
\ee
Here $\delta G_{T = 0}$ is the temperature-independent contribution that results from a small interaction-induced change in the Fermi level.
To calculate the relative correction from the interaction, $\delta G_H - \delta G_F$ should be  divided by (\ref{G0}) (we are interested only in the temperature-dependent term and $g_s=2$) to give
\be
\frac{\delta G_{T}}{G_0} = \nu_2\,[U_p(0) - 2 U_p(2k_F)]\,\frac{T}{E_F},
\ee
where $\nu_2 = {m}/({\pi \hbar^2})$ is the spinful density of states.
The relative correction linearly depends on temperature and is much larger than the standard Fermi-liquid $T^2$ corrections. It is a consequence of the cusp in a transmission coefficient (Fig. \ref{fig3}). This is essentially the same temperature dependence that was obtained previously for the correction to the conductance of wide ballistic contacts and was due to the collisions of the injected and incident electrons in the leads\cite{Nagaev08}.

The absolute value of the temperature-dependent correction to the conductance is proportional to the fourth power of the contact size $\delta G \propto (k_F a)^4$ and the relative one $\delta G/G_0$ does not depend on $a$. This is in contrast with the correction for wide contacts $\delta G_{semi}$, which is proportional to
$(k_F a)^2\ln(l_c/a)$, where
$l_c \gg a$ is a large cutoff length due to a scattering by impurities or a finite size of the sample, so that $\delta G/G_0|_{semi}$ is roughly proportional to $G_0$. Should this correction be extrapolated to narrow contacts,
it would be proportional to $G_0^2$ because both the number of injected and incident electrons is proportional to $G_0$. Hence the correction from Friedel oscillations must dominate at small contact widths.

The quantum correction from the Friedel oscillations is more sensitive to the shape of interaction potential than
the semiclassical one.
In particular, its sign is determined by the factor $[U_p(0) - 2 U_p(2k_F)]$, which is due to a competition between a positive contribution from the exchange interaction and a negative contribution from the direct interaction. This factor arises in several theories of scattering by Friedel oscillations\cite{Zala, Rudin, Sablikov} and is positive for long-range and negative for short-range interactions.

\begin{figure}[t]
 \includegraphics[width=0.6\linewidth]{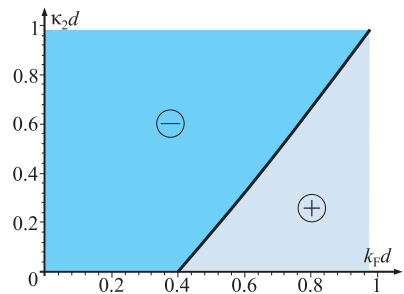}
 \caption{\label{fig4} The sign of the correction resulting from Friedel oscillations vs. $k_F d$ and
 $\kappa_2 d$. The grey region corresponds to positive sign, and the blue, to negative.
 }
\end{figure}

Consider the most typical case of the Coulomb potential
statically screened by a metallic gate parallel to the 2D
electron gas and the electrons in the gas itself. If the
distance between the gate is $d$ and the dielectric constant
of environment is $\varepsilon_d$, the interaction potential between the
electrons in the gas is given by (see Appendix C)
\be
U(q) = \frac{4 \pi e^2}{\varepsilon_d (\coth|qd| + 1) |q| + 4\pi e^2 \nu_2},
\ee
which leads to the correction of the form
\begin{multline}
\frac{\delta G_{T}}{G_0} = \frac{T}{E_F}\,\Biggl[\frac{2\kappa_2 d}{1 + 2\kappa_2 d}
\\
- \frac{2\kappa_2}{k_F[\coth(2k_F d) + 1] + \kappa_2}\Biggr],
\label{dG/G_0-2}
\end{multline}
where $\kappa_2$ is the inverse screening length. Figure 4 shows the regions in
the $(k_F d,\kappa_2 d)$ plane where the correction is positive or negative.

Unfortunately, we are unaware of detailed measurements of the temperature dependence of the contact conductance near pinch-off.
The estimation for realistic values of distance to the gate  $d = 100 \, {\rm nm}$, the inverse screening length  $\kappa_2 = 2\pi e^2 \nu_2 / \varepsilon_d = 1.93 \times 10^6 \, {\rm cm}^{-1}$ and electron density $n_s = 10^{-10} \, {\rm cm}^{-2}$ for $T = 1K$ results in $\delta G_T/G_0 = - 15\%$.
Measurements of conductance in the region of the 0.7-anomaly, indeed, reveal a negative slope of $G(T)$ dependence\cite{Cronenwett}. However a quantitative comparison with our predictions is not possible.

In the multichannel regime, this negative correction should be suppressed by the positive one from the scattering of oppositely moving electrons\cite{Nagaev08}, and one should observe a change in the sign of the slope of its temperature dependence.
However if the actual $\kappa_2$ is smaller due to the  low electron concentration, the sign of the correction from the electron-electron interaction may remain positive for all contact sizes. Therefore one may estimate the actual screening length from the sign and slope of the temperature dependence of conductance for narrow contacts.

Though our results were obtained for a sharp potential in a form of an infinitely narrow and very high barrier, they should also survive for reasonably smooth potentials. It is easily seen  that the smoothness of potential results only in a phase shift of the Friedel oscillations and preserves the  $x^{-3/2}$ dependence of their amplitude far from the barrier (see Appendix B). That's why the linear temperature dependence and change of sign of slope of $G(T)$ should be robust with respect to the exact shape of the barrier.

In summary, we calculated the conductance of a narrow and short quantum point contact at nonzero temperature taking into account the electron-electron interaction. The conductance linearly depends on temperature and is proportional to the fourth power of the contact size, and the relative correction does not depend on the contact size. The sign of the linear temperature-dependent term depends on the competition between direct and exchange interaction. Measurements of the slope of its temperature dependence allow one to determine the parameters of electron-electron interaction.

\begin{acknowledgments}
This work was supported by Russian Foundation for Basic Research, grants 10-02-00814-a and 11-02-12094-ofi-m-2011, by the program of Russian Academy of Sciences, the Dynasty Foundation, and by the Ministry of Education and Science of Russian Federation, contract  No 16.513.11.3066.
\end{acknowledgments}

\appendix
\section{Calculation of transmitted wave function}

We use the approach described by Sommerfeld\cite{Sommerfeld} for the problem of two-dimensional diffraction and solve the system (\ref{chi-system}) assuming a continuity of the derivative of the wave function $\psi = \psi_0 + \psi_t$ at $x=0$, where $\psi_0$ is the wave function in the absence of the gap and $\psi_t$ is the correction in the lowest order in a contact size.
We expand the unknown function into a Fourier integral
\be
\psi_t(x, y) = \int\limits_{-\infty}^{\infty}{\frac{dk_y}{2\pi}\,e^{-i k_y y}\, \psi_t(x, k_y)}
\label{FT}
\ee
and then search a solution in the form of outgoing waves $\psi_t(x, k_y) = c_1 \times e^{- i k_x x}$.
The constant $c_1$ is easily expressed from the boundary condition
$$
 c_1 = \int\limits_{-\infty}^{\infty}{dy'\, e^{i k_y y'}\, \chi(0, y')}.
$$
We sequentially substitute these formulas into (\ref{FT}) and obtain
\be
\psi_t(x, y) = \int\limits_{-\infty}^{\infty}{dy' \, \chi(0, y') \, K(x, y, y')},
\label{psi_t}
\ee
where the kernel is given by
\be
K(x, y, y') = -\frac{i}{2}\, k x \, \frac{H_1^{(1)}(k\sqrt{x^2 + (y - y')^2})}{\sqrt{x^2 + (y - y')^2}}.
\label{K}
\ee

The condition of continuity of the derivative of total wave function at the gap may be written as
\begin{multline}
\frac{\partial \psi_0(x, y)}{\partial x}\big|_{x = -0} + \frac{\partial \psi_t(x, y)}{\partial x}\big|_{x = -0} = \frac{\partial \psi_t(x, y)}{\partial x}\big|_{x = +0},
\label{deriv_cont}
\end{multline}
where
$$
 \psi_0(x, y) = \sqrt{\frac{m}{k_x}} \, e^{i k_y y} \, (e^{i k_x x} - e^{- i k_x x}).
$$
We substitute the kernel (\ref{K}) into (\ref{deriv_cont}) and obtain an integral equation
\be
\int\limits_{-a}^{a}{dy' \, \chi(0, y') \, \frac{k H_1^{(1)}(k |y - y'|)}{|y - y'|}} = 2 \sqrt{\frac{m}{k_x}} \, k_x \, e^{i k_y y}
\ee
This integral equation appears in a problem of a two-dimensional diffraction by a narrow slit\cite{Sommerfeld}.
We are interesting in the case of $k \approx k_F$, so the limit $k_F a \ll 1$ is equivalent to $ka \ll 1$. The solution of the integral equation in the limit of $ka \ll 1$ is given by\cite{Polyanin}
\be
\chi(0, y') = -i \, k_x \, \sqrt{\frac{m}{k_x}}\, \sqrt{a^2 - y'^2}
\label{chi}
\ee
We substitute (\ref{chi}) and (\ref{K}) into (\ref{psi_t}) and obtain
\be
\psi_t(x, y) = \frac{\pi}{4} \, k_x k a^2 \, |x| \, \sqrt{\frac{m}{k_x}} \, \frac{H_1^{(1)}(k r)}{r}.
\ee
It is conveniently presented in the form
\be
\psi_t(x, y) = -\frac{i\pi}{2}\sqrt{\frac{m}{k_x}}a^2 k_x
\int\limits_{-\infty}^{\infty}{\frac{dq_y}{2\pi} e^{i (q_x x + q_y y)}}.
\label{fourier_psi}
\ee

In a similar way we solve the system (\ref{g-eq}) for the Green's function  of the Schr\"odinger
equation $g_t$
and obtain
\be
g_t = \frac{\pi}{8} \, \frac{m}{\hbar^2} \, k^2 a^2 \, |x||x'| \, \frac{H^{(1)}_1(k r) H^{(1)}_1(k r')}{r r'}.
\ee

\section{Friedel oscillations of electron density}

The oscillating electron density is given by
\begin{multline}
 n({\bf r}, {\bf r_1}) = \langle \hat\psi^{+}(\br_1)\,\hat\psi(\br)\rangle
 =
 \sum_{\alpha} f(\eps_{\alpha})\,\psi_{\alpha}^{*}(\br_1)\,\psi_{\alpha}(\br),
 \label{n}
\end{multline}
where $\psi_{\alpha}$ are the eigenfunctions of the non-interacting Hamiltonian.
In the lowest  approximation in the contact size we neglect the distortion of the Friedel oscillations of electron density due to the gap and consider them as arising from a solid barrier. Because of the translational symmetry in the direction parallel to the barrier we may use instead of $\alpha$  momentum $p$  and the eigenfunctions
\be
\psi_p(\br) = e^{i p_y y}[e^{i p_x x} + r(p_x)e^{- i p_x x}]
\label{psi_rp_x}
\ee
With the help of distribution function $f(p)$, Eq. (\ref{n}) may be rewritten in the form
\be
n({\bf r}, {\bf r_1}) = \int{\frac{d^2 p}{(2\pi)^2}\, f(p)\, \psi_p^*({\bf r_1})\, \psi_p({\bf r})}.
\label{n(r', r3)}
\ee
In the case of a sharp infinitely high barrier, the reflection coefficient $r(p_x)=-1$ for any $p_x$ and the substitution
of $\psi_p(x, y)$ into (\ref{n(r', r3)}) gives us after simple rearrangements

\be
n({\bf r}, {\bf r_1}) = 4\int{\frac{d^2 p}{(2\pi)^2}}\,\cos[p_y (y - y_1)]
\nonumber\\
\times
\sin(p_x x)\,\sin(p_x x_1)
\ee
Then we go to cylindrical coordinates $(p, \p)$, expand the product of trigonometric functions into a sum of four terms and perform the integration over $\p$.  The remaining integral over $\p$ is a sum of four integrals of the form
\be
\int\limits_{-\pi/2}^{\pi/2}{d\p \, \cos(a\cos\p + b\sin\p)} = \pi \, J_0(\sqrt{a^2 + b^2}).
\ee
Calculations give the electron density in the form
\begin{multline}
n({\bf r}, {\bf r_1}) = \int\limits_{0}^{\infty}{\frac{dp}{2\pi}}\, f(p)\, p \left[J_0(p\sqrt{(x - x_1)^2 + (y - y_1)^2})
\right.
\nonumber\\
\left.
-
J_0(p\sqrt{(x + x_1)^2 + (y - y_1)^2})\right].
\end{multline}

Upon integrating  by parts, one arrives at
\begin{multline}
n({\bf r}, {\bf r_1}) = \frac{1}{2\pi}\int\limits_{0}^{\infty}{dp \left(-\frac{\partial f}{\partial p}\right)}
\\{}\times
p
\left[\frac{J_1(p\sqrt{(x - x_1)^2 + (y - y_1)^2})}{\sqrt{(x - x_1)^2 + (y - y_1)^2}}\right.
\\-
\left. \frac{J_1(p\sqrt{(x + x_1)^2 + (y - y_1)^2})}{\sqrt{(x + x_1)^2 + (y - y_1)^2}}\right].
\end{multline}
In the case of $\br_1 = \br$ the Friedel oscillations of density depend only on $x$, so
\begin{multline}
n({\bf r}, {\bf r}) = n(x) =  \frac{1}{2\pi}\int\limits_{0}^{\infty}{dp \left(-\frac{\partial f}{\partial p}\right)}
\\{}\times
\left[\frac{p^2}{2} - \frac{pJ_1(2px)}{2x}\right].
\label{n(r,r)-sharp}
\end{multline}

In the case of a smooth yet impenetrable barrier the wave function $\psi_p$ away from it still may be presented in the form (\ref{psi_rp_x}) with a momentum-dependent reflection coefficient $r(p_x)= -\exp[i\delta(p_x)]$, where $\delta(p_x)$ presents the phase shift of the reflected wave with respect to the case of zero boundary conditions at $x=0$. We substitute (\ref{psi_rp_x}) into the Eq. (\ref{n(r', r3)}) and consider the coordinate-dependent oscillating part of electron density.
With the help of cylindrical coordinates $(p, \p)$ we can write it in the form
\be
n_{osc}(\br, \br) = - \frac{1}{\pi^2}\int\limits_{0}^{\infty}{dp\, f(p)\,p \int\limits_{0}^{\pi/2}{ d\p }}
\nonumber\\ \times
\cos(2px\cos\p + \delta(p\cos\p)).
\ee
At large distances from the barrier $2px \gg 1$ we can estimate $n_{osc}(\br, \br)$ using the stationary phase method \cite{}
\be
n_{osc}(\br, \br) = - \frac{1}{2\pi^2}\int\limits_{0}^{\infty}{dp\, f(p)\, p}
\nonumber\\ \times
\left[\sqrt{\frac{\pi}{px} } \cos(2px - \pi/4 + \delta(p))
+ {\cal O} \left(\frac{1}{(2px)^{3/2}}\right)\right]
\ee
If $\partial \delta(p)/\partial p \ll 2x$, it is possible to integrate this equation by parts and to obtain
\begin{multline}
n_{osc}(\br, \br) = \frac{1}{4(\pi x)^{3/2}}\int\limits_{0}^{\infty} dp \left(-\frac{\partial f}{\partial p}\right)
\\{}\times
p^{1/2}\cos(2px + \delta(p) + \pi/4).
\label{n(r,r)-smooth}
\end{multline}
It is easily seen that at low temperatures and far from the contact, $n_{osc}$ exhibits  the same asymptotic behavior  as the oscillating part of Eq. (\ref{n(r,r)-sharp}) except for the phase shift $\delta(p_F)$.

Elucidate now the conditions for validity of (\ref{n(r,r)-smooth}).
If the screening in the system is two-dimensional and the potential of the barrier falls off according to a power law, $\partial \delta(p)/\partial p \sim x_0(p)$, where $x_0(p)$ is the classical turning point for the electrons at the barrier. Even if the width of the barrier $2x_0(p_F)$ is of the order of the Fermi wavelength, there is a large interval of distances $p_F^{-1} \ll x \ll v_F/T$ where Eq. (\ref{n(r,r)-smooth}) holds, and it is precisely this interval that dominates the temperature-dependent contribution  from Friedel oscillations to the conductance.

\section{Screened Coulomb potential}

\begin{figure}[t]
 \includegraphics[width=0.6\linewidth]{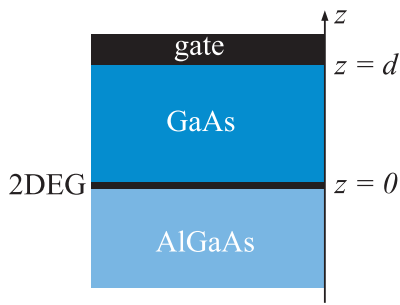}
 \caption{\label{fig5} A model of experimental structure.}
\end{figure}

We calculate Coulomb potential screened by a gate and two-dimensional electrons. Consider the system shown in
Fig. \ref{fig5} with a positively charged particle $e_0$ at point $(0, 0, 0)$. The total potential induced by the charged particle, the two-dimensional electrons and the gate satisfies the Poisson equation
\be
- \nabla^2 \phi(\br_{||}, z) = \frac{4\pi}{\varepsilon}[\rho^{ext} + \rho^{ind}],
\label{poisson}
\ee
where $\rho^{ext} = e_0 \delta(\br_{||})\delta(z)$ is the density of the particle charge and $\rho^{ind} = - e^2 \nu_2 \delta(z)\phi(\br)$ is the density of the induced charge calculated in Thomas-Fermi approximation. We take a Fourier transform of (\ref{poisson}) with respect to the in-plane coordinates
and integrate it with respect to $z$ over a small vicinity of $z=0$ to arrive at an equation
\be
\frac{\partial \phi(q, z)}{\partial z}\big|_{z= - 0}^{z = + 0} - 2\kappa_2\phi(q, 0) = - \frac{4 \pi e_0}{\varepsilon}.
\label{d phi}
\ee
Here $\kappa_2 = 2/a_B$ is the inverse two-dimensional screening length. We assume that the layer of 2DEG is thin and the potential is continuous at $z = 0$, i.e.
\be
\phi(q, z= - 0) = \phi(q, z = + 0).
\label{phi}
\ee
We solve Eqs. (\ref{d phi}) and (\ref{phi}) with boundary conditions
\be
&\phi(q, d) = 0\\
&\phi(q, -\infty) = 0
\ee
and write the potential in the form
\be
\phi(q, 0) = \frac{4\pi e_0}{\varepsilon}\frac{1}{[\coth(q d) + 1]q + 2\kappa_2}.
\label{phi_q}
\ee
The potential in the coordinate space is obtained by inverse Fourier transform of Eq. (\ref{phi_q}) and may be conveniently expressed in terms of a dimensionless coordinate $x = qr_{||}$
\be
\phi(\br_{||}, 0) = \frac{2e_0}{\varepsilon r_{||}}\int{dx\frac{J_0(x)}{\coth(x d/r_{||}) + 1
+ 2\kappa_2 r_{||}/x}}.
\label{four}
\ee
In experiments\cite{Renard, Khrapai} the case of $d \gg \kappa_2^{-1}$ is realized. An evaluation of the integral (\ref{four}) gives for different limiting cases
\be
\phi(\br_{||}, 0) = \left\{
\begin{aligned}
\frac{e_0}{2\varepsilon \kappa_2^2 r^3_{||}}, \qquad r_{||} \gg d \gg \kappa_2^{-1}\\
\frac{e_0}{\varepsilon \kappa_2^2 r^3_{||}}, \qquad d \gg r_{||} \gg \kappa_2^{-1} \\
\frac{e_0}{\varepsilon r_{||}}, \qquad d \gg \kappa_2^{-1} \gg r_{||}.
\end{aligned}
\right.
\ee


\bigskip\qquad

\end{document}